# Universal valence-band picture of

# the ferromagnetic semiconductor GaMnAs

Shinobu Ohya[*], Kenta Takata, and Masaaki Tanaka[†]

*Department of Electrical Engineering and Information Systems, The University of Tokyo, 7-3-1 Hongo, Bunkyo-ku, Tokyo 113-8656, Japan*

*e-mail: ohya@cryst.t.u-tokyo.ac.jp

[†]e-mail: masaaki@ee.t.u-tokyo.ac.jp

Abstract

The origin of ferromagnetism in the prototype ferromagnetic semiconductor GaMnAs is still controversial due to the insufficient understanding of its band structure and Fermi level position.  Here, we show the universal valence-band (VB) picture of GaMnAs obtained by resonant tunneling spectroscopy for a variety of surface GaMnAs layers with the Mn concentrations from 6 to 15% and the Curie temperatures from 71 to 154 K.  We find that the Fermi level exists in the bandgap, and that the VB structure of GaAs is almost perfectly maintained in all the GaMnAs samples, *i.e.* VB is *not* merged with the impurity band.  Furthermore, the *p-d* exchange splitting of VB is found to be quite small (only several meV) even in GaMnAs with a high Curie temperature (154 K). These results indicate that the VB structure of GaMnAs is quite insensitive to the Mn doping.

Extensive studies on the prototype ferromagnetic semiconductor GaMnAs in the last decade have revealed a variety of unique features induced by the combination of its magnetic and semiconducting properties. The attractive features and functionalities, such as the electric-field control of the Curie temperature ($T_C$)[1,2,3] and of the magnetization direction[4], and the magnetization reversal induced by a quite low current density[5,6] or with the spin-orbit interaction[7], are expected to be utilized for future ultra low-power spintronic devices. In spite of these studies, however, the basic physics lying in this material including the mechanism of the ferromagnetism and the band structure has not been fully understood yet[8]. The mean-field Zener model, where the ferromagnetism is induced by the *p-d* exchange interaction between the valence-band (VB) holes up to $10^{20}$-$10^{21}$ cm$^{-3}$ and the localized Mn-3*d* electrons, has been generally accepted until recent years, because it seems to be able to explain a variety of features of GaMnAs[9,10,11,12,13]. In this model, it is assumed that the VB is merged with the Mn induced impurity-band (IB). The Fermi level position is determined by the concentration of the VB holes, and the Fermi level lies at 200-300 meV lower than the top of the merged VB as shown in Fig. 1(a). (Here, we refer to this model as the VB conduction model.) On the other hand, several reports on the optical[14,15,16,17] and transport[18,19] properties of GaMnAs have shown that $E_F$ exists in IB within the bandgap of GaMnAs. In this case, the Zener double-exchange-type mechanism is applicable for GaMnAs, where the ferromagnetism is stabilized by hopping of the spin-polarized holes in IB[20]. Meanwhile, a recent scanning tunneling microscope analysis has shown the microscopically inhomogeneous nature of the electronic structure of GaMnAs, making the understanding of this material more difficult[21]. This controversial situation stems from the insufficient understanding on the VB structure of GaMnAs.

Resonant tunneling spectroscopy is a very powerful method for clarifying the band structure of ferromagnetic semiconductors. The resonant levels contain abundance of material parameters, such as the Fermi level, effective mass (*i.e.* band dispersion), band offset, and strain. Therefore, this method gives us quite useful clues to the mechanism of the ferromagnetism as well as the precise band picture. We proved the effectiveness of this method for a limited case; the GaMnAs quantum wells (QWs) *sandwiched* by double barriers[22,23], where the control of the ferromagnetism of the GaMnAs QW was difficult because the interstitial Mn defects confined by the double barriers cannot be easily removed out. Here, we find that this method can be applied also for the surface *non-embedded* GaMnAs layers for the first time. This new finding allows us to systematically investigate a variety of GaMnAs films with a wide range of the Mn contents and $T_C$ values, where the characteristics of the surface GaMnAs can be easily controlled by low temperature annealing[24]. In this paper, using this method and precise etching technique for various GaMnAs surface layers, we clarify the universal VB picture of GaMnAs [Fig. 1(b)].

To investigate the VB structures of GaMnAs, we have used the 200-μmϕ mesa diodes composed of $Ga_{1-x}Mn_xAs$ (*d* nm)/ AlAs (5 nm)/ GaAs:Be (100 nm, Be concentration: $1\times10^{18}$ $cm^{-3}$) grown on $p^+$GaAs(001) substrates, where *d* was precisely controlled by changing the etching time from mesa to mesa [Fig. 2(a)]. Figure 2(b) depicts the ideal example of the behavior of the resonant levels formed in the surface GaMnAs layer when *d* is increased from $d_1$ to $d_2$ ($d_1 < d_2$). The black solid curves, red dash-dotted lines and blue lines express the VB top energy ($E_V$) when the quantum-size effect is neglected, the Fermi level and the resonant levels, respectively. The strain and exchange split are neglected for simplicity. These resonant levels are formed by the holes confined by the AlAs barrier and the surface Schottky barrier induced by Fermi-level pinning at the surface. These resonant levels can be detected only when the holes are injected from the GaAs:Be layer because the injected holes have a small in-plane wave vectors $k_\parallel$ due to the small hole concentration ($1\times10^{18}$ $cm^{-3}$) of the GaAs:Be electrode[22]. In our analyses, energy bands with the symmetry of the p-orbitals are mainly detected in GaMnAs, reflecting the holes' feature injected from the GaAs:Be electrode. Ideally, these resonant levels in the surface GaMnAs layer converge at $E_V$ in the bulk GaMnAs with increasing *d* [Fig. 2(c)]. Table 1 shows the sample parameters and characteristics investigated here. In the growth of Sample D with *x* of 15%, it was difficult to control the precise growth condition due to its high Mn concentration[25], which is the reason for the thin film thickness of Sample D: 4.6 nm. The bias voltage *V* is defined by the voltage of the top electrode with respect to the substrate.

Figure 3(a)-(d) show the $d^2I/dV^2$-*V* curves obtained for various *d* at 3.4 K in Sample A-D, respectively. In all the samples, oscillations due to resonant tunneling are clearly detected. They tend to be clearer with increasing $T_C$ (*i.e.* from Sample A to D), which means that the coherency of the tunneling holes is enhanced with increasing $T_C$ regardless of their high Mn concentrations. Here, HH*n* and LH*n* (*n*=1, 2, 3, . . .) are resonant tunneling through the *n*th level of the heavy-hole (HH) band and light-hole (LH) band in GaMnAs assigned by the theoretical analyses mentioned below, respectively. The LH1 and HH2 resonant peaks are overlapped in Sample A-C due to the compressive strain in GaMnAs as explained later. The HH1 level is observed in all the samples shown in Fig. 3(a)-(d), indicating that the HH1 level is not occupied by holes in the equilibrium condition. Since the HH1 is the highest VB resonant level and the Fermi level position corresponds to the zero bias, our results clearly indicate that the Fermi level lies in the bandgap in all the samples. These resonant oscillations persist up to 30-60 K (See Supplementary Information A and Fig. S1.), which contrasts with the weak coherent feature of the holes near the Fermi level reported by Wagner *et al.* at low temperatures less than 1 $K^{26}$. These results mean that the feature of the wave functions of the VB hole is largely different from the one at the Fermi level. In Sample A and B, the resonant levels tend to converge at ~-0.12 V and at ~-0.09 V with increasing *d*, respectively [Fig. 3(a) and (b)]. Therefore, we can see that the Fermi

level becomes closer to the VB top position with increasing $T_C$ or the hole concentration. If we assumed that the fluctuation of the energy band structure existed in the in-plane area of these mesa diodes with 200 μm in diameter, such clear resonant peaks would not be observed. Therefore, the VB structure is spatially *homogeneous* unlike the *inhomogeneous* electronic structure of the Mn-induced impurity states formed at various energy levels recently observed by the microscopic scanning tunnel microscopy measurements[21]. In Sample C, we see systematic shifts of the resonant peaks with increasing $d$ when $d$ is less than 15 nm [Fig. 3(c)]. When $d$ gets larger than 15 nm, however, opposite shifts of the resonant peaks to the larger voltages are observed with increasing $d$. This is because the VB is slightly modulated by the interstitial Mn atoms diffused to the surface during the annealing[24]. This scenario is supported by the theoretical analyses mentioned below.

We calculated the quantum levels in these heterostructures by the multiband transfer matrix technique developed in Ref. 27 with the Luttinger-Kohn 6×6 *k·p* Hamiltonian and strain[10]. (For details, see the Methods section.) For reproducing the experimental data of the resonant peak bias voltages, we assumed the VB diagram shown in Fig. 4(a) for Sample A, B and D, and assumed the one shown in Fig. 4(b) for Sample C. Here, the black solid lines correspond to $E_V$ when the strain and *p-d* exchange split are neglected. We divided the surface Schottky barrier region, whose width is estimated to be less than 1 nm by the Poisson equation, into two [Fig. 4(b)] or three [Fig. 4(a)] equally-spaced regions and assumed that the band is flat in each region. For other regions, the small band bending effect was neglected, and flat bands were assumed. For Sample C, we assumed the existence of a shallow QW region in the surface GaMnAs layer as shown in Fig. 4(b) in order to reproduce the anomalous behavior of the peaks' shifts observed when $d$ is larger than 15 nm [Fig. 3(c)]. This QW region is considered to be formed by the diffusion of interstitial Mn atoms (acting as donors) toward the surface during the annealing. Figure 4(c) shows the schematic VB structures of GaAs:Be and GaMnAs as a function of the in-plane wave vector $k_\parallel$ when the quantum-size effect is neglected. $E_F$ is defined by the energy difference between the Fermi level and HH band at the Γ point in GaMnAs. Note that the HH and LH bands are split due to the compressive strain in GaMnAs. In the following, the value of *s* corresponds to the coefficient of the relationship between the bias voltage -*V* and the energy relative to the Fermi level[28,29]. We used $E_F$, the deformation anisotropic term $Q_\varepsilon$ and *s* value as fitting parameters. The values of $d$ of the calculated data were corrected so that the experimental data were well reproduced, where we define *r* as the ratio of the etching rate estimated by fitting to the experimental data as mentioned above to that estimated experimentally (See the Methods section).

The color-coded $d^2I/dV^2$ intensities of Sample A-D are shown in Fig. 5(a), (b), upper graph of (c), and the one of (d), respectively. The calculated resonant-peak bias voltages of the HH and LH bands as a function of $d$ are expressed by the connected violet and green dots, respectively.  For Sample A and B [Fig. 5(a) and (b)], we fit the resonant peaks by assuming $s = 1.4$, which means that the 40% of the voltage drop at the AlAs barrier is consumed in other areas including the surface Schottky barrier.  In the case of Sample C, we changed $s$ gradually with increasing $d$ as shown in the lower graph of Fig. 5(c).  Also, the resistance area ($RA$) as a function of $d$ when $V$ is -100 mV is shown in this graph, where we can see roughly the similar tendency between $RA$-$d$ and $s$-$d$.  The increase in $RA$ in the large $d$ region ($d$ >15 nm) indicates that an additional voltage drop occurs near the surface, supporting our assumption of the increase in $s$ in the large $d$ region.  This result can be attributed to the diffusion of the interstitial Mn atoms toward the surface during the annealing.  Due to the large $x$ (9%) in Sample C, this effect influences the VB position more remarkably than in Sample A and B [See Fig. 4(b)].  Note that $RA$ is presumably affected also by the density of states of the Mn-induced IB around the Fermi level, which is considered to be the major factor preventing the perfect correspondence between $RA$-$d$ and $s$-$d$.  In the case of Sample D, we also gradually changed the $s$ value with increasing $d$ as shown in the lower graph of Fig. 5(d).  The $s$ values obtained in this study are consistent with those obtained in our previous studies on the double barrier heterostructures with a GaMnAs QW[22,23] (See Supplementary Information B for details).  As can be seen above, the resonant peaks observed in all of our samples are well reproduced by our model, which gives the evidence that the resonant levels are formed by the quantization of the VB holes.

Figure 5(e) shows the $E_F$ values obtained above as a function of the hole concentration estimated by using other literatures on GaMnAs films with the Mn concentrations, $T_C$ values and thicknesses similar to our films[11,13].  (See Supplementary Information C and Fig. S2 for details on the measurements for Sample E with $x$ of 6%.) We see that the $E_F$ systematically decreases with increasing the hole concentration. This tendency is consistent with the results of the infrared optical absorption measurements reported by Burch *et al*[14].  The difference of the $E_F$ values between the results in Ref. 14 (100-200 meV) and our results (50-80 meV) is probably due to the fact that the exact determination of the Fermi level by the optical measurement is a little difficult due to the broad spectrum from IB, while our method detects the Fermi level position more directly.  (For more comments on the $E_F$ position, see Supplementary Information D.)

As can be seen above, we can well describe the observed resonant levels by our model using the band parameters of *GaAs* (See the Methods section for details).  This result means that the VB features of GaAs including the effective masses of holes are almost perfectly maintained and that VB is *not* merged with IB in GaMnAs, as shown in Fig. 1(b)[8].  Also, other important result is that the *p-d* exchange split of the resonant levels is not observed at all.  The *p-d* exchange split has been thought to be very strongly associated with the appearance of the ferromagnetism in GaMnAs[30,31].  If we assumed the value of $N_0\beta$ to be 1.0 eV[32], the exchange splitting energy of the LH band $xN_0\beta<S>/3$ would be 50 meV for $x=6\%$, where $N_0$, $\beta$ and $<S>$ are the cation density, *p-d* exchange and the thermal average of magnetic spin, respectively.  This splitting energy would further increase with increasing *x*.  However, we do not see such split peaks in any samples investigated here.  Therefore, we conclude that the *p-d* exchange split is negligibly small (several meV by our previous spin-dependent resonant tunneling study[23]) in GaMnAs.

Our results clearly indicate that the Fermi level lies in the bandgap, which means that the carrier transport occurs in the bandgap and suggests that the Zener double-exchange-type mechanism is likely to be applicable for GaMnAs[20].  However, there have been no reports on the clear detection of IB in the energy region around several tens meV higher than the VB top in the bandgap.  The solid curves in Fig. 6(a)-(d) show the $d^2I/dV^2$-$V$ characteristics in the low bias region (-$V$<0.1 V) for various *d* in Sample A-D at 3.4 K, respectively.  We see that the several peaks, which are traced with the broken curves in Fig. 6(b)-(d), gradually appear and become clearer with increasing $T_C$ (*i.e.* from Sample A to D).  These peaks do not largely depend on *d*, which means that the effective mass of these states is very large.  They cannot be explained only by assuming the magnon or the LO phonon usually discussed in such a low bias region in the GaMnAs heterostructures[33].  Therefore, they might correspond to a part of IB induced by the Mn doping, although more detailed analysis is necessary.

In summary, we have observed the resonant tunneling levels formed by the quantization of VB in the surface GaMnAs layers with the various Mn concentrations (6-15%) and $T_C$ values (71-154 K) for the first time.  These resonant levels are well explained by assuming the effective masses of holes in *GaAs*, which means that the VB structure of GaMnAs is almost the same as that of GaAs and that VB is *not* merged with IB.  We have found that $E_F$ exists in the bandgap in all the GaMnAs films, which suggests that the Zener double-exchange-type mechanism is applicable for GaMnAs.  No exchange splitting of these resonant levels has been observed, which indicates that the *p-d* exchange interaction is negligibly small (less than several meV).

**Acknowledgments**
This work was partly supported by Grant-in-Aids for Scientific Research, the Special Coordination Programs for Promoting Science and Technology, FIRST Program by JSPS, PRESTO of JST and Asahi Glass Foundation.

**Methods**
**- Sample preparations and transport measurements**
The investigated heterostructures comprising $Ga_{1-x}Mn_xAs$(*d* nm)/ AlAs (5 nm)/ GaAs:Be (100 nm, Be concentration: $1\times10^{18}$ cm$^{-3}$) were grown by molecular beam epitaxy on p$^+$type GaAs(001) substrates.  The sample parameters and characteristics are described in Table 1. The GaAs:Be and AlAs layers were grown at 600 and 550ºC, respectively. The growth temperatures of the surface GaMnAs layer were 215, 215, 210 and 185ºC for Sample A, B, C and D, respectively. Sample A was an as-grown sample. The Sample B, C and D were annealed at 160, 160, 140ºC for 45, 34, and 95 hours after the growth, respectively. The circular mesa diodes with 200 μm in diameter were fabricated by chemical etching. $d$ was controlled by changing the etching time from mesa to mesa, where $d$ was estimated from the etching rate obtained by using a surface profiler. We spin-coated an insulating resist on the sample, opened a contact hole with 180 μm in diameter on the top of the mesas, and fabricated a metal electrode by evaporating Au on this surface.

Tunneling transport measurements were carried out in a cryostat cooled at 3.4 K with a conventional two-terminal direct-current (DC) method. The $d^2I/dV^2$-$V$ characteristics were numerically derived by differentiating the data of the $I$-$V$ characteristics.

**- Calculations**

For calculating the quantum levels in the surface GaMnAs layer, we used the multiband transfer matrix technique developed in Ref. 27 with the Luttinger-Kohn 6×6 $k\cdot p$ Hamiltonian and strain Hamiltonian[10], where the coherent tunneling of the VB holes was assumed. Because the exchange splitting was too small to be detected in our measurements, we neglected the $p$-$d$ exchange interaction[10] in our calculation. The Luttinger parameters $\gamma_1$, $\gamma_2$ and $\gamma_3$, split-off energy gap $\Delta_{so}$ and deformation potential $b_V$ of GaMnAs were assumed to be the same as those of GaAs; 6.85, 2.1, 2.9, 0.34 eV and -1.9 eV, respectively. For calculating the quantum levels, we assumed that the in-plane wave vector $k_\parallel$ is 0 because of the small Fermi surface of the holes in the GaAs:Be electrode.

**References**

1. H. Ohno *et al.*, Nature **408,** 944-946 (2000).
2. I. Stolichnov *et al.,* Nature mater. **7,** 464-467 (2008).
3. M. Sawicki *et al.*, Nature Phys. **6,** 22 (2009).
4. D. Chiba *et al.*, Nature **455,** 515-518 (2008).
5. D. Chiba *et al.*, Phys. Rev. Lett. **93,** 216602 (2004).
6. M. Watanabe *et al.*, Appl. Phys. Lett. **92,** 082506 (2008).
7. A. Chernyshov *et al.*, Nature phys. **5**, 656-659 (2010).
8. T. Jungwirth *et al.,* Phys. Rev. B **76,** 125206 (2007).
9. T. Dietl et al., Science **287**, 1019-1022 (2000).
10. T. Dietl, H. Ohno, and F. Matsukura, Phys. Rev. B **63,** 195205 (2001).
11. T. Jungwirth, *et al.,* Phys. Rev. B **72,** 165204 (2005).
12. D. Neumaier *et al.*, Phys. Rev. Lett. **103,** 087203 (2009).
13. Y. Nishitani *et al.*, Phys. Rev. B **81,** 045208 (2010).
14. K. Hirakawa *et al.*, Phys. Rev. B **65**, 193312 (2002).
15. K. S. Burch *et al.,* Phys. Rev. Lett. **97,** 087208 (2006).
16. V. F. Sapega *et al.*, Phys. Rev. Lett. **94,** 137401 (2005).
17. K. Ando *et al.*, Phys. Rev. Lett. **100,** 067204 (2008).
18. L. P. Rokhinson *et al.*, Phys. Rev. B **76,** 161201(R) (2007).
19. K. Alberi *et al.,* Phys. Rev. B **78,** 075201 (2008).
20. H. Akai, Phys. Rev. Lett. **81,** 3002-3005 (1998).
21. A. Richardella *et al.*, Science **327,** 665 (2010).
22. S. Ohya *et al.*, Phys. Rev. B **75,** 155328 (2007).
23. S. Ohya *et al.*, Phys. Rev. Lett. **104,** 167204 (2010).
24. K.W. Edmonds *et al.,* Phys. Rev. Lett. **92,** 037201 (2004).
25. S. Ohya *et al.*, Appl. Phys. Lett. **90**, 112503 (2007).
26. K. Wagner *et al.*, Phys. Rev. Lett. **97,** 056803 (2006).
27. A. G. Petukhov *et al.*, Phys. Rev. Lett. **89,** 107205 (2002).
28. H. Ohno *et al.*, Appl. Phys. Lett. **73,** 363-365 (1998).
29. M. Elsen *et al.,* Phys. Rev. Lett. **99,** 127203 (2007).
30. K. Sato *et al.,* Reviews of Modern Physics **82**, 1633 (2010).
31. P. Mahadevan and A. Zunger, Phys. Rev. B **69,** 115211 (2004).
32. J. Okabayashi *et al.*, Phys. Rev. B **59,** R2486 (1999).
33. H. Saito *et al.*, J. Appl. Phys. **103,** 07D127 (2008).

**Figure captions**

**Figure 1** (a) Band structure of GaMnAs when assuming the VB conduction model. VB is merged with the Mn-induced impurity band (IB) formed in the bandgap, which results in the change of the effective masses of holes. This model contradicts our results. (b) VB picture of GaMnAs obtained in our study. VB is not largely affected by the Mn doping in GaMnAs. We find that VB of GaMnAs is reproduced by that of GaAs (i.e. with the same effective mass of holes in GaAs) with strain and a small $p$-$d$ exchange splitting (3-5 meV). Although we do not directly detect IB in this study, IB is likely to exist in the bandgap.

**Figure 2** (a) Schematic device structure investigated in this study composed of GaMnAs($d$ nm)/ AlAs(5 nm)/ GaAs:Be (Be: $1\times10^{18}$ cm$^{-3}$, 100 nm)/ p$^+$GaAs(001) junctions. (b) Schematic band diagrams of these heterostructures when $d=d_1$ and $d=d_2$ ($d_1 < d_2$). The black solid curves, red dash-dotted lines and blue lines express the VB top energy ($E_V$) when the quantum-size effect is neglected, Fermi level and the resonant levels of the VB holes, respectively. The strain and exchange splitting are neglected for simplicity. These resonant levels are formed by the holes confined by the AlAs barrier and the surface Schottky barrier induced by Fermi level pinning at the surface. (c) Ideal $d$ dependence of the resonant energies. The converged energy corresponds to $E_V$ in the bulk GaMnAs.

**Figure 3** (a)-(d) The solid curves correspond to the $d^2I/dV^2$-$V$ curves for the various GaMnAs thicknesses $d$ of Sample A- D, respectively. The broken curves are the traces of the resonant peaks. The assignments of these peaks are carried out based on the results of our calculation of the resonant levels. Note that the axis scale of $d$ shown in (d) is different from those in (a)-(c) due to the small initial thickness of the GaMnAs layer in Sample D. The color in these graphs expresses the $d^2I/dV^2$ intensity extrapolated from the measured data. All the measurements were carried out at 3.4 K.

**Figure 4** The VB diagram assumed for Sample A, B and D is shown in (a), and the one assumed for Sample C is shown in (b). The black solid lines correspond to $E_V$ when the strain and $p$-$d$ exchange splitting are neglected. The red dash-dotted lines correspond to the Fermi level. In (b), we assumed a shallow QW with a steep VB wall at $z = d$-10 nm (when $d > 10.9$ nm) or at 0.9 nm (when $d \leq 10.9$ nm) in the surface GaMnAs layer. In the real heterostructure, this is not a steep wall but a slope induced

by the gradual change of the interstitial Mn concentration in the growth direction. (c) The schematic VB structures of GaAs:Be and GaMnAs as a function of the in-plane wave vector $k_\parallel$ when the quantum size effect is neglected. $E_F$ is defined as the energy difference between the Fermi level and the HH energy at the Γ point in GaMnAs. Note that the HH and LH bands are split due to the compressive strain in GaMnAs.

**Figure 5** (a), (b), upper graph of (c), upper graph of (d) The color-coded $d^2I/dV^2$ intensities of Sample A, B, C and D, respectively. The white dots located at the top of these graphs correspond to the $d$ values of the devices used for the measurements, where the data at these dots correspond to those shown in Fig. 3(a)-(d). Other region is drawn by interpolation. The connected violet and green dots are the calculated resonant peak bias voltages of the HH and LH bands, respectively. The values of $d$ of the calculated data were corrected so that the experimental data were well reproduced, where we define $r$ as the ratio of the etching rate obtained by fitting to the experimental data to that estimated experimentally (See the Methods section). Lower graphs of (c) and (d) The assumed $s$ values in the calculation and the experimentally obtained $RA$ values as a function of $d$ at 3.4 K. (e) The $E_F$ values obtained in our study as a function of the hole concentration estimated by using the literatures for GaMnAs films with the Mn concentration, $T_C$, and thickness similar to our films[11,13]. The characters in this graph correspond to the sample names. (For details of Sample E, see Supplementary Information C.)

**Figure 6** (a)-(d) The solid curves correspond to the $d^2I/dV^2$-$V$ curves in the small bias (-$V$ < 0.1 V) region for the various GaMnAs thicknesses $d$ of Sample A-D, respectively. Note that the axis scale of $d$ shown in (d) is different from those of others due to the small initial thickness of the GaMnAs layer in Sample D. The color in these graphs expresses the $d^2I/dV^2$ intensity extrapolated from the measured data. All the measurements were carried out at 3.4 K.

**Table 1** Sample parameters, characteristics and estimated energies ($E_0$ and $E_F$) of the GaMnAs-based devices investigated in this study. Here, $a$ (nm) is the initial GaMnAs thickness before etched. The $T_C$ values were estimated by the Arrott plot using the data of the magnetic field dependence of magnetic circular dichroism measured for these samples before etched. See Fig. 4(a)-(c) for the definitions of $E_0$ and $E_F$. For details of the measurements on Sample E, see the Supplementary Information C.

| Sample | $x$ (%) | $a$ (nm) | $T_C$ (K) | $E_0$ (meV) | $E_F$ (meV) |
|--------|---------|----------|-----------|-------------|-------------|
| A      | 6       | 21       | 71        | -89         | 80          |
| B      | 6       | 21       | 111       | -69         | 60          |
| C      | 9       | 22       | 146       | -87         | 52          |
| D      | 15      | 4.6      | 154       | -77         | 59          |
| (E)    | 6       | 21       | 131       | -61         | 52          |

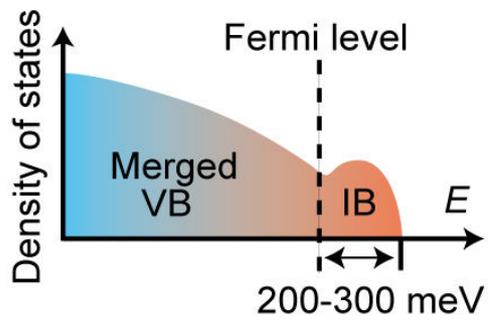 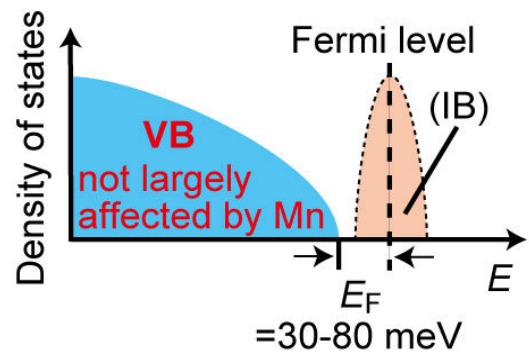

Figure 1

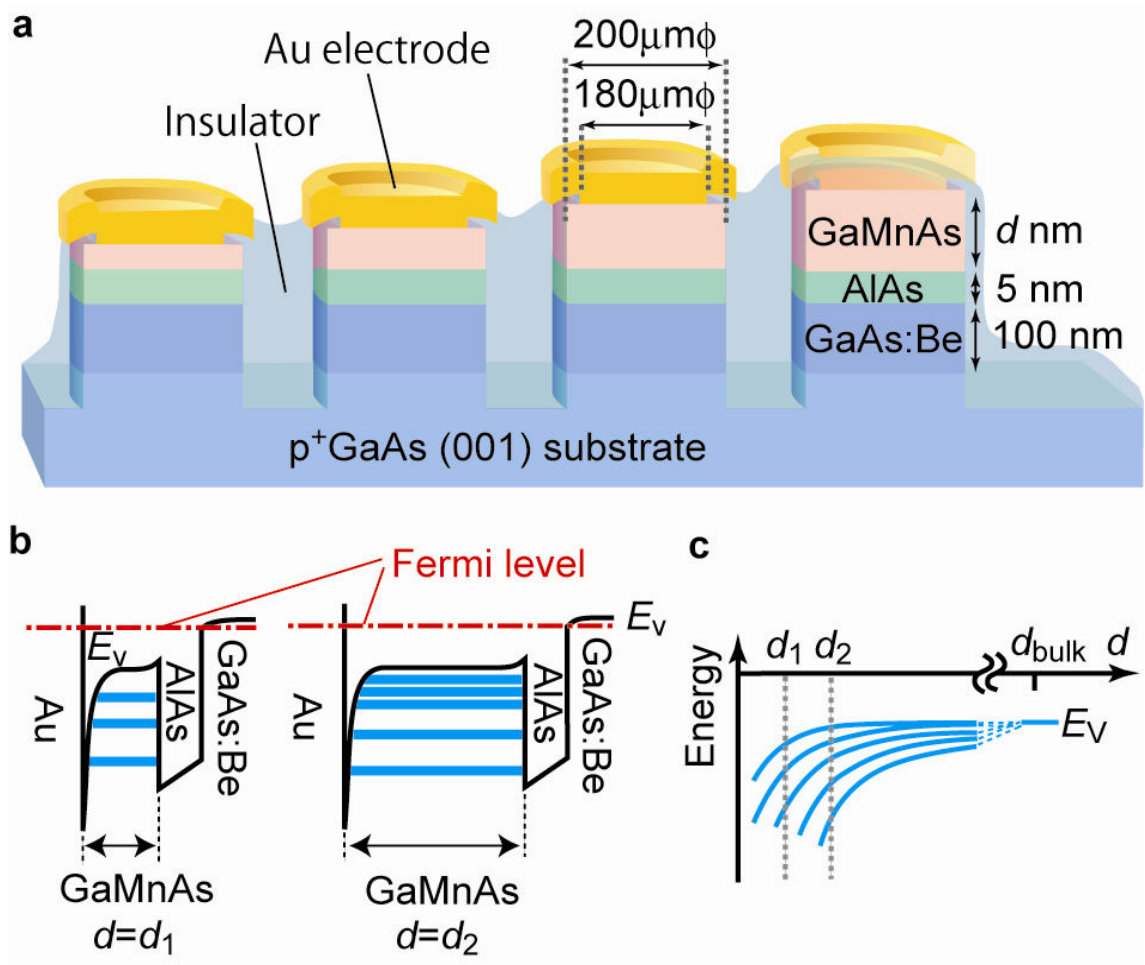

Figure 2

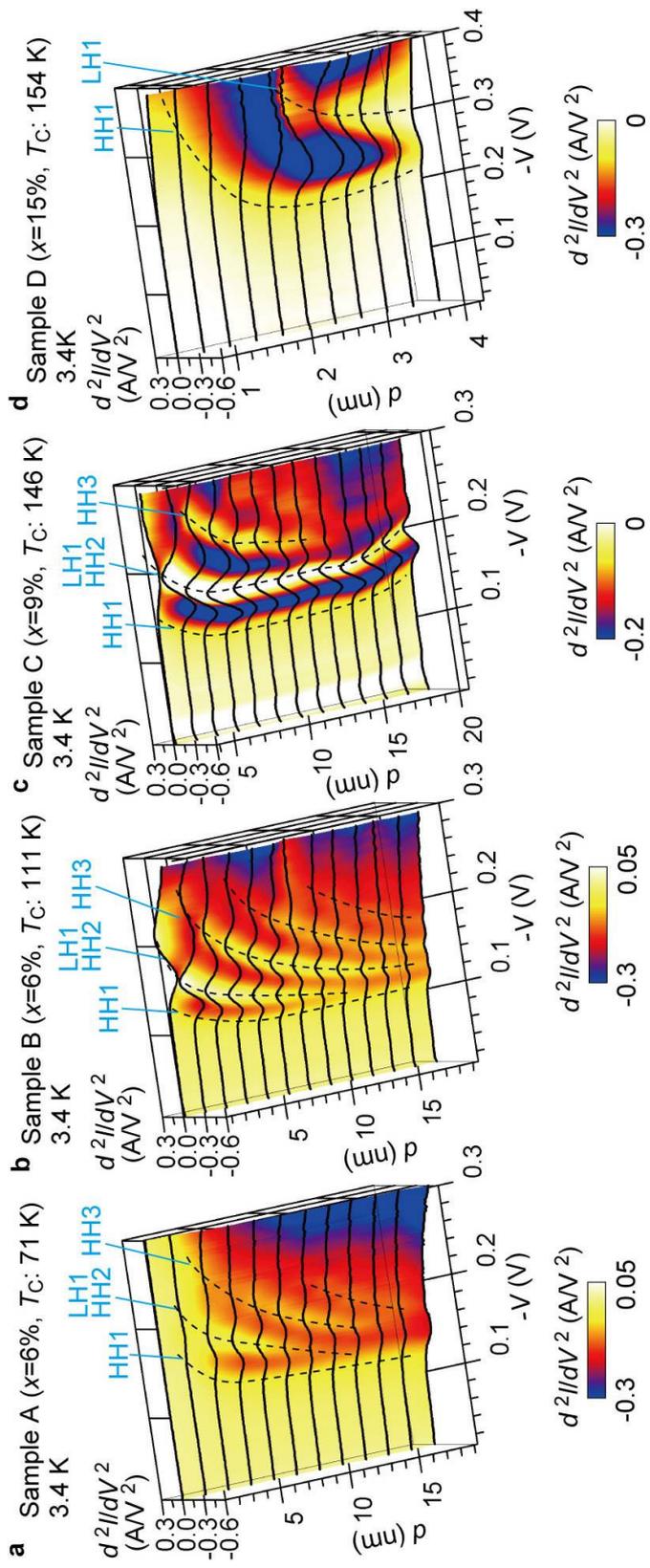

Figure 3

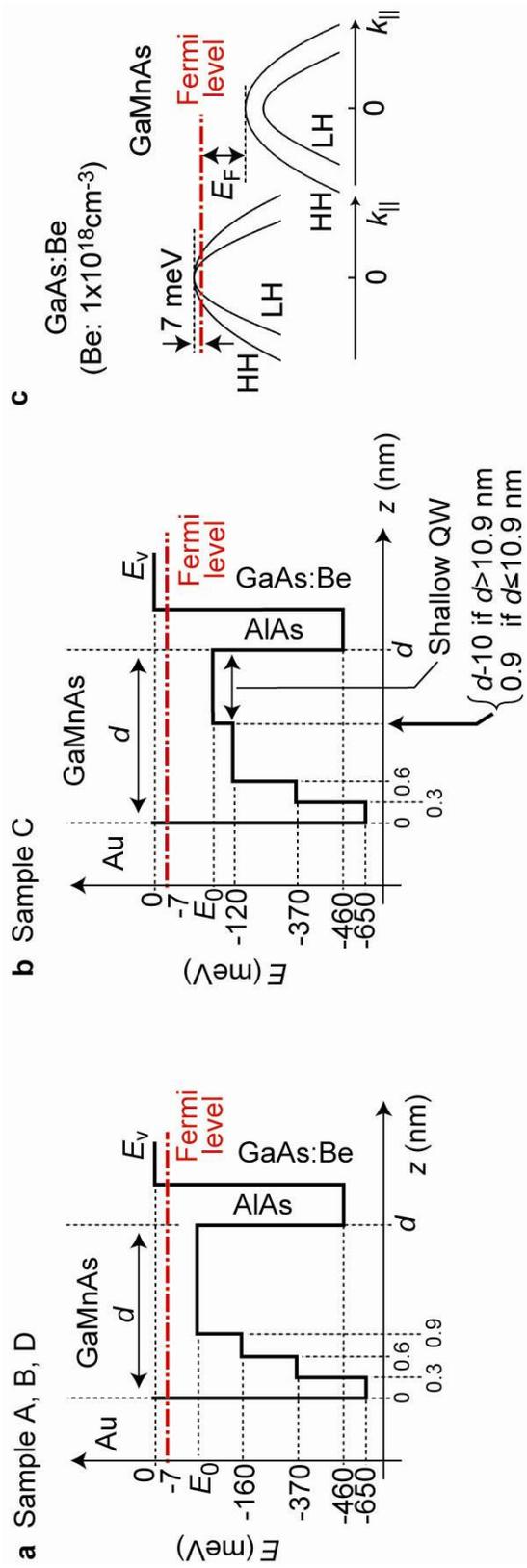

Figure 4

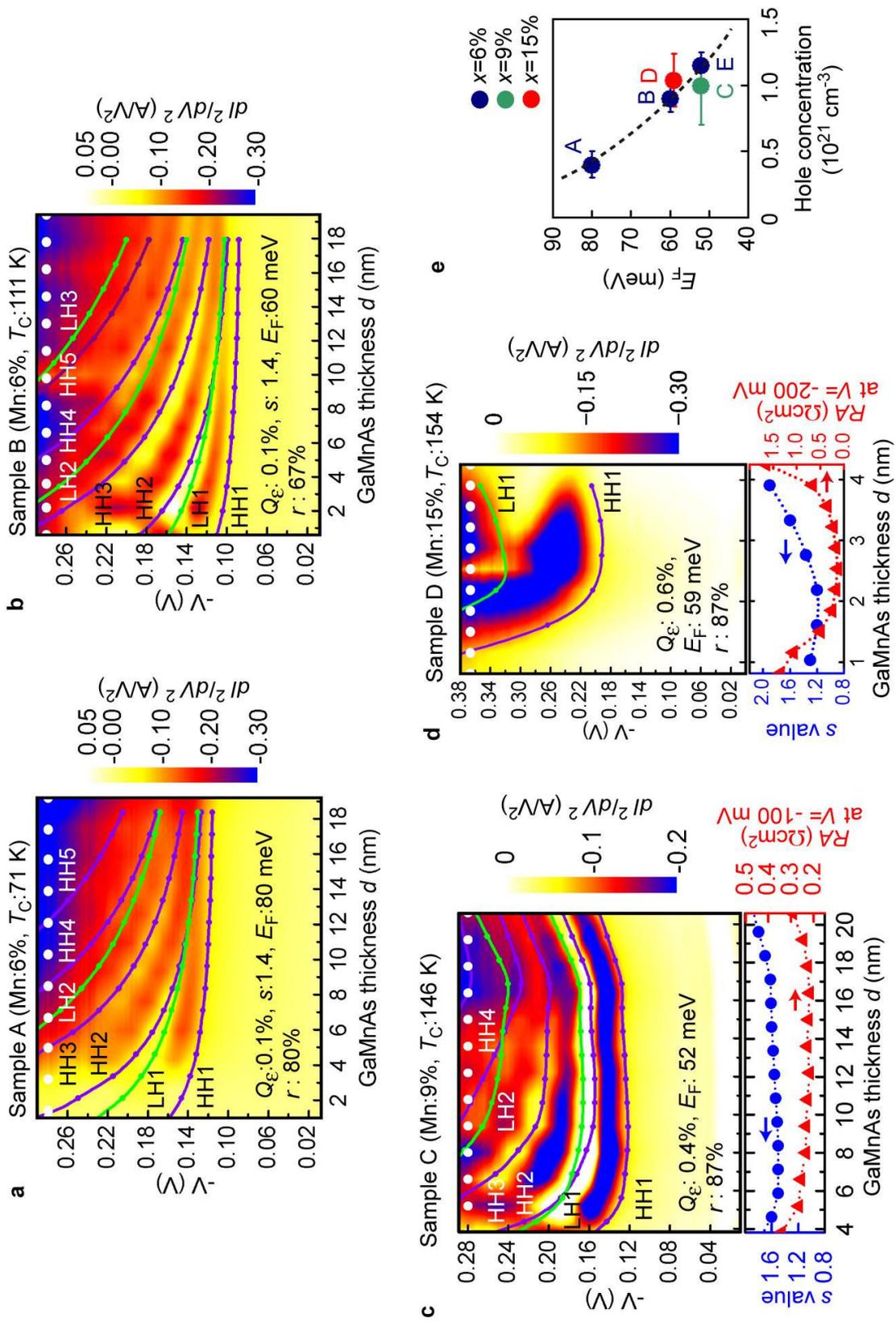

Figure 5

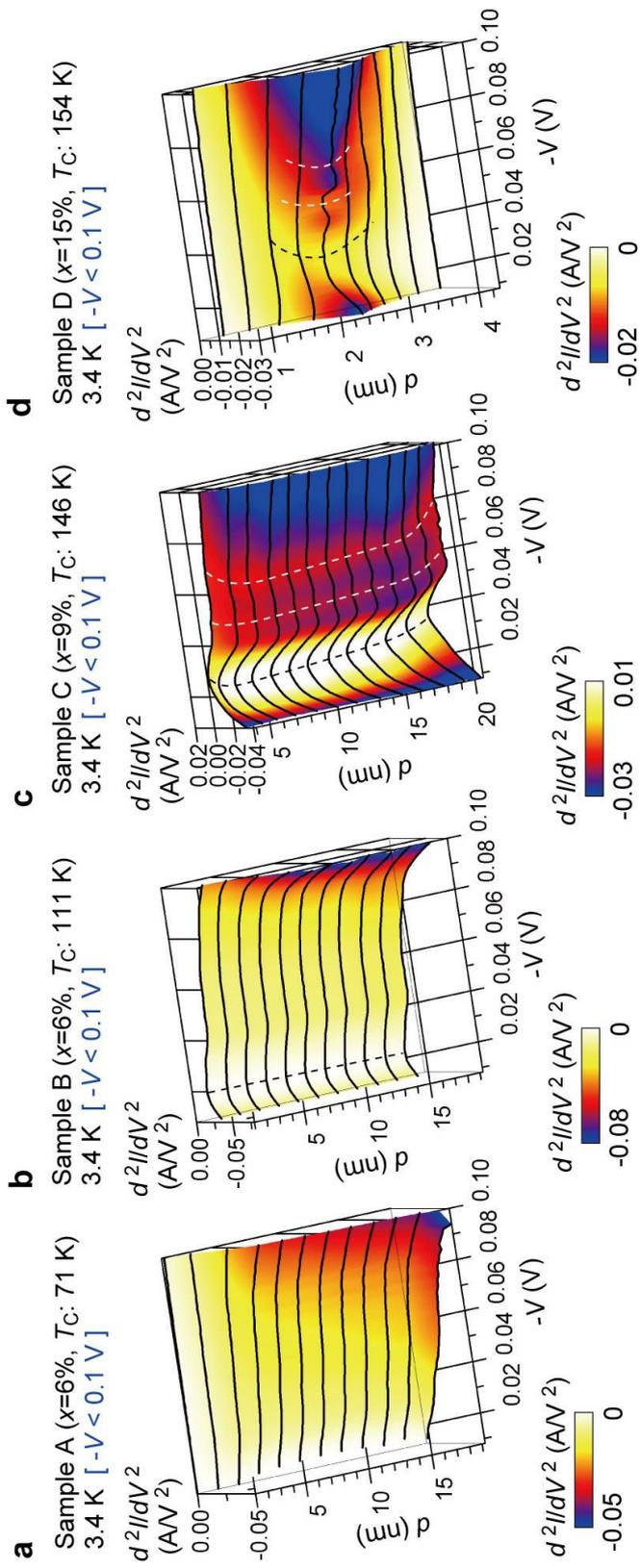

Figure 6

## Supplementary Information

A. Temperature dependence

We measured the temperature evolution of the $d^2I/dV^2$-$V$ characteristics of the sample composed of $Ga_{0.91}Mn_{0.09}As$ (13.6 nm)/ AlAs(5 nm)/ GaAs:Be(100 nm) grown on a $p^+$GaAs(001) substrate. The growth conditions of this sample are the same as those of Sample C shown in Table 1 in our manuscript. This sample was annealed at 160ºC for 12 hours after the growth, and the Curie temperature was 139 K. The solid curves shown in Fig. S1 correspond to the $d^2I/dV^2$-$V$ curves measured at various temperatures. We see that the resonant peaks persist up to 30-60 K.

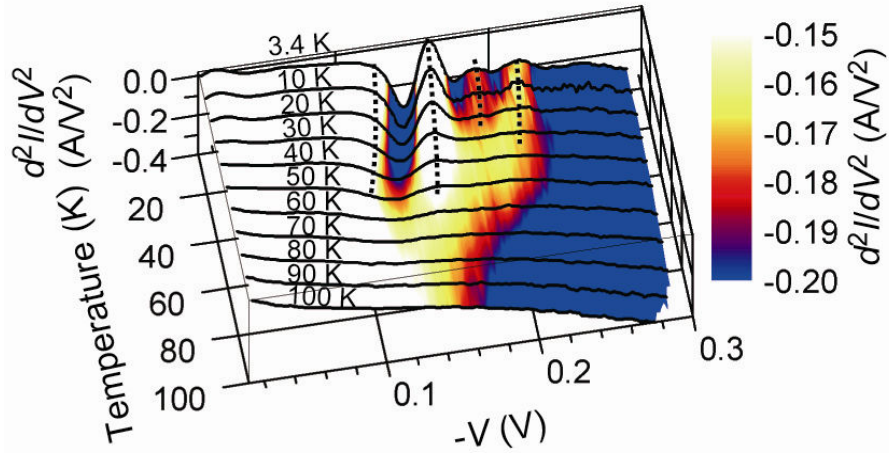

Figure S1  Temperature evolution of the $d^2I/dV^2$-$V$ characteristics obtained in the heterostructure composed of $Ga_{0.91}Mn_{0.09}As$ (13.6 nm)/ AlAs(5 nm)/ GaAs:Be(100 nm) grown on a $p^+$GaAs(001) substrate.  The color expresses the $d^2I/dV^2$ intensity extrapolated from the measured data.

## B. Validity of the *s* values obtained in this study

In the double-barrier GaMnAs-QW heterostructures in our previous studies[1,2], the *s* parameters were estimated to be from 2.4 to 3. Thus, the voltage drop in other area than the double barriers in these heterostructures is estimated to be from 0.4 (= 2.4 - 2) to 1 (= 3 - 2) times that consumed at one barrier. Therefore, the *s* values obtained in the surface GaMnAs layer studied here (1.2 - 2) are consistent with our previous studies.

## C. Measurements on the Sample E (Mn 6%, $T_C$: 131 K)

Sample E is composed of $Ga_{0.94}Mn_{0.06}As$ (*d* nm)/ AlAs(5 nm)/ GaAs:Be(100 nm) grown on a $p^+$GaAs(001) substrate. The initial thickness of GaMnAs was 21 nm before etched. By the annealing at 160ºC for 46 hours after the growth, a high $T_C$ value of 131 K was obtained in this sample. Figure S2 shows the color-coded $d^2I/dV^2$ intensities of Sample E as functions of –*V* and *d* at 3.4 K. The white dots located at the top of the panel correspond to the *d* values of the devices used for the measurements. Other region is drawn by interpolation. We could obtain only 4 mesas from this sample due to mishandling in the device preparation. The connected violet and green dots are the calculated resonant peak bias voltages of the HH and LH bands when assuming the VB structure shown in Fig. 4a in the main text, respectively. Here, $E_0$, $E_F$ and $Q_\varepsilon$ were assumed to be -61 meV, 52 meV and 0.1%, respectively.

## D. About the fluctuation of $E_F$ due to the difference of the growth conditions

We note that the $E_F$ position seems to fluctuate due to the difference of the growth conditions. Our previous study on the $Ga_{0.94}Mn_{0.06}As$-QW double-barrier heterostructures, where the GaMnAs QW was grown at 225-240ºC, showed that $E_F$ was only ~30 meV in spite of its low $T_C$ ~30 K[22]. Since it is very plausible to think that the density of states of IB is influenced by the substitutional Mn concentration that is determined by the growth conditions[3], it is likely to influence the IB formation and thus the Fermi level position.

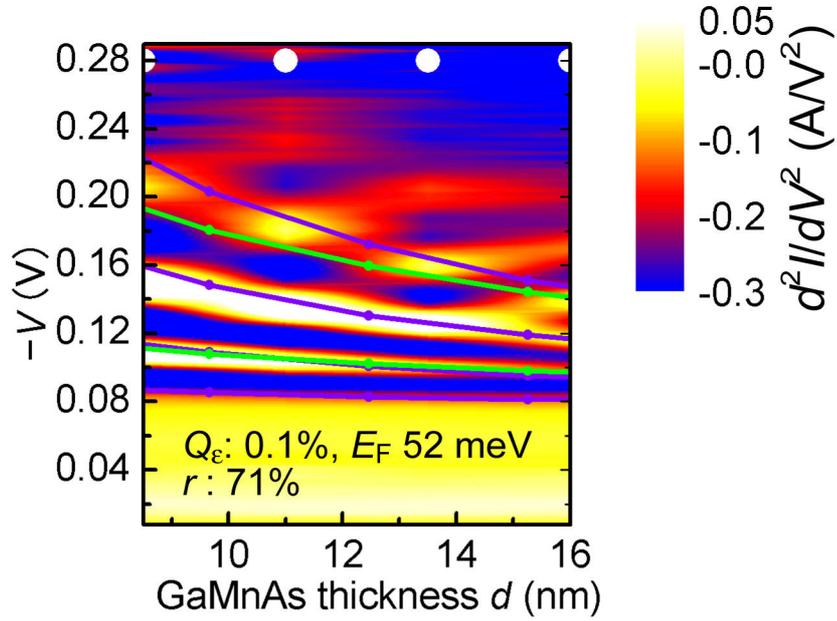

Figure S2  The color-coded $d^2I/dV^2$ intensity of Sample E as functions of $-V$ and $d$. The white dots located at the top of the graph correspond to the $d$ values of the devices used for the measurements. Other region is drawn by interpolation. The connected violet and green dots are the calculated resonant peak bias voltages of the HH and LH bands, respectively.

References in Supplementary Information

[1] S. Ohya, P. N. Hai, Y. Mizuno, and M. Tanaka, Phys. Rev. B **75,** 155328 (2007).

[2] S. Ohya, I. Muneta, P. N. Hai, and M. Tanaka, Phys. Rev. Lett. **104,** 167204 (2010).

[3] S. Ohya, K. Ohno, and M. Tanaka, Appl. Phys. Lett. **90,** 112503 (2007).